\documentclass[a4paper,twocolumn,prb]{revtex4}

\usepackage{graphicx}
\usepackage{dcolumn}
\usepackage{bm}

\def\dfrac#1#2 {{\displaystyle {#1 \over #2}}}
\def\dint {\mathop{\displaystyle \int}}

\def\dsum {\mathop{\displaystyle \sum}}

\def\dfrac#1#2 {{\displaystyle {#1 \over #2}}}
\def\dint {\mathop{\displaystyle \int}}

\def\dsum {\mathop{\displaystyle \sum}}
\begin{document}

\title{A statistical approximation to solve ordinary differential equations}

\author{M. Febbo}
\email{mfebbo@uns.edu.ar}
\author{M. L. Alemany}
\author{S. A. Vera}
\affiliation{Departamento de F\'{\i}sica, Universidad Nacional del
Sur, Av. Alem 1253, 8000, Bah\'{\i}a Blanca, Argentina}
\date{\today}

\begin{abstract}

We propose a physical analogy between finding the solution of an
ordinary differential equation (ODE) and a $N$ particle problem in
statistical mechanics. It uses the fact that the solution of an
ODE is equivalent to obtain the minimum of a functional. Then, we
link these two notions,  proposing this functional to be the
interaction potential energy or thermodynamic potential of an
equivalent particle problem. Therefore, solving this statistical
mechanics problem amounts to solve the ODE. If only one solution
exists, our method provides the unique solution of the ODE. In
case we treat an eigenvalue equation, where infinite solutions
exist, we obtain the absolute minimum of the corresponding
functional or fundamental mode. As a result, it is possible to
establish a general relationship between statistical mechanics and
ODEs which allows not only to solve them  from a physical
perspective but also to obtain all relevant thermodynamical
equilibrium variables of that particle system related to the
differential equation.

\end{abstract}
\maketitle

\section{INTRODUCTION}
Ordinary differential equations are one way physicists model
physical phenomena. To solve them, a great variety of analytical
and numerical methods are available in the literature
\cite{Hildebrand73,Kaplan85}.

Trough this article we wish to solve some problems that arise in
vibration theory making an analogy of them with a system of
particles, a well known issue treated in Statistical Mechanics. If
this analogy is made, all the tools used in Statistical Mechanics
are available for tackling these type of problems. For example,
one is able to calculate the entropy or the specific heat of the
system, something which is, at least, difficult to imagine in a
typical vibration problem. From a more general perspective, we
extend the same analogy and apply it to find the solutions of
ordinary differential equations (ODEs) of general  type (see
section \ref{sec:Mathematicalp}).

The general method proposed here could be resumed in the following
way. From a mathematical viewpoint, it is possible to obtain the
solution of an ODE from the minimization of a functional.  On the
other hand, a thermodynamic potential is represented by a
functional. We connect these two notions,  proposing the
functional which leads to the solution of an ODE, to be the
interaction potential energy or thermodynamic potential of an
equivalent particle problem. Therefore, solving this statistical
mechanics problem amounts to solve the ODE. For the case where
only one solution exists, the method allows to obtain the unique
solution of the ODE. Clearly, when we treat vibrational problems,
an eigenvalue equation must be solved with infinite possible
solutions. This case our method is capable of obtaining the
fundamental mode or absolute minimum of the corresponding
functional.

The principal aim of our work is to show (a) how the previously
proposed correspondence is used to attain the solution of ordinary
differential equations, (b) the theoretical basis upon which it is
sustained and (c) the way it can be treated within a stochastic
framework.

The present work is organized as follows. In section
\ref{sec:Mathematicalp}, we show how the problem of solving a
general kind of ODE can be transformed to a problem of
minimization of a functional (weak solution of the differential
equation \cite{Rektorys75}). Section \ref{sec:Statisticalt}
presents the foundations of equilibrium statistical analysis
viewed as a minimization of a thermodynamic potential. This
provides the theoretical the basis of our developed method.
Section \ref{sec:Proposeda} describes the general aspects of the
numerical scheme and algorithm implementation. Section
\ref{sec:Langevina} is devoted to find the solution of an ODE
employing a stochastic framework. This method allow us to solve
the ODE proposing an equivalent Langevin (stochastic) equation. At
the end, in section \ref{sec:Numericals}, several numerical
experiments which will sustain our theoretical hypothesis will be
developed and solved. Finally, some conclusions will be outlined.

\section{MATHEMATICAL PROBLEM}\label{sec:Mathematicalp}
There are many possible approaches to find the solution of ODEs
\cite{Hildebrand73,Kaplan85}. From all of them, we are interested
in that one which transforms the problem of finding the solution
of an ODE into the minimization of a functional. To do this we
must recall some results of functional analysis.

\subsection{Obtaining a minimum principle for the differential equation}
The most general form of an ODE of order $2k$ can be written as
\cite{Rektorys75}
\begin{equation}\label{eq:ODE}
    Au(x)=\dsum_{i=0}^{k}(-1)^i (p_i u^{(i)}(x))^{(i)}=f
\end{equation}
where $u(x)$ is an element of a Hilbert space $\mathcal{H}_A$,  $x
\in \Re$ and $u^{(i)}(x)$ is the derivative of order $(i)$ of
$u(x)$. Also $f \in L_2(a,b)$ ($L_2(G)$ Hilbert space of functions
square integrable in the domain $G$) and $p_i(x),i=0,1,...,k$ are
functions continuous with their derivatives up to the $i-th$ order
inclusive in the closed interval $[a,b]$ which satisfy
\begin{equation}\label{eq:pi}
    p_i(x)\geq 0 \ \text{in} [a,b], \ i=0,1,...,k-1
\end{equation}
\begin{equation}\label{eq:pik}
    p_k(x)\geq p \ \text{in} [a,b], \ p=const.
\end{equation}
For the rest of the paper, homogeneous boundary conditions will be
considered with no loss of generality.

Then, it is possible to look for the generalized solution
$u^{\ast}$ of equation (\ref{eq:ODE}) as  that element which
minimizes the functional
\begin{equation}\label{eq:funct}
    F(u)=\dsum_{i=0}^{k} \dint_{0}^{L} p_i (u^{(i)})^2 dx
    -2\dint_{0}^{L}fu dx
\end{equation}

in the corresponding Hilbert space $\mathcal{H}_A$, where operator
$A$ is defined. There exists a requirement of positive
definiteness of operator $A$ that ensures that the extreme value
of (\ref{eq:funct}) or (\ref{eq:eigen-funct}) is the desired
solution $u^{\ast}$ of the problem\footnote{It can be proved that
the solution exists and is unique but the demonstration is beyond
the scope of this paper}.

For the eigenvalue problem, $Au-\lambda u=0, \ \lambda \in \Re$,
the functional to be minimized coincides with the corresponding
eigenvalue $\lambda$ and it reduces to
\begin{equation}\label{eq:eigen-funct}
    F(u)\equiv\lambda=\frac{\dsum_{i=1}^{k} \dint_{0}^{L} p_i (u^{(i)})^2 dx}{\dint_{0}^{L}u^2 dx}
\end{equation}
In this case, infinite minima exist, each of them corresponding to
the associated eigenvalues
\begin{equation}\label{eq:eigenvalues}
    \lambda_1\leq \lambda_2\leq ...
\end{equation}
Equation (\ref{eq:eigenvalues}) states that the first (or
fundamental) eigenvalue $\lambda_1$ is the minimum possible
eigenvalue and the corresponding eigenfunction is $u_1$; i. e.
$F(u_1)$ is the absolute minimum of $F(u)$ \cite{Rektorys75}.

Summing up, a differential operator can be transformed into a
functional so that minimizing this functional equals to solving
the ODE.

\section{STATISTICAL TREATMENT}\label{sec:Statisticalt}

As a second step towards attaining the solution of ODEs,  we must
adequate the problem to be treated as a standard particle problem
in statistical mechanics. To this end, we must think on a $n$
particle system in a one dimensional box of size $L$, where a
point in the $n$ dimensional configurational phase space can be
denoted by $\mathbf{q}=\{q_1,q_2,...q_n\}$, being $q_j$ the
position of particle $j$. Now, a link between the representation
of this system in configurational phase space $\mathbf{q}$ with
the description of possible candidate solutions of the ODE $u(x)$
(see equation \ref{eq:ODE}) is proposed through an adequate
discretization of $u(x)$.

\subsection{Discretization of $u(x)$}

Consider now any possible candidate to the solution of a
particular kind of ODE, named $u(x)$. This continuous function is
an element in the corresponding Hilbert space $\mathcal{H}_A, x
\in \Re$ where operator $A$ is defined, and belongs to the domain
of the functional $F(u)$. The proposed discretization of $u$ will
be settled from the consideration of $n$ sites labelled with the
discrete index $l$ of a 1 dimensional lattice. Then, the function
$u(x)$ is taken as the limit $n \rightarrow \infty$ from the field
functions $u_l(x_l)$. So the link between our system of particles
and that element $u(x)$ is naturally settled as $u_l(x_l)=q_l$.
Each $u_l$ can take any value between $-\infty$ and $\infty$. A
possible configuration of our $n$ particle system is then
considered by specifying the value of $u_l$ at each site $l$ or
simply the lattice vector
$\mathbf{q}=\{u_1(x_1),...,u_l(x_l),...u_n(x_n)\}$. Through this
connection, it is possible to apply the whole
statistical framework to treat this equivalent new problem.\\

\subsection{Equilibrium statistical analysis}

The next step is to formulate the problem as in standard
statistical mechanics. To do this, we will consider our particle
system in the canonical ensemble. So far we have made a connection
between the domain of the functional $F(u)$ with the microstates
of a thermal system, represented by $\mathbf{q}$. With this in
mind, and using the expression for the probability density in the
canonical ensemble $\rho_{\mathbf{q}}=e^{-\beta H(\mathbf{q})}/Z$,
our proposal is to define an equivalent problem, such that the
Hamiltonian $H(\mathbf{q})$ is replaced by our functional
$F(\mathbf{q})$ i. e.
\begin{equation}\label{prop}
\rho_{\mathbf{q}}= \frac{\exp{(-\beta F(\mathbf{q}))}}{Z}
\end{equation}
Here, as well as in standard statistical mechanics $Z$ represents
the partition function ( this time $Z=\int e^{-\beta
F(\mathbf{q})} d\mathbf{q}$) and $\beta$ can be considered a
parameter equivalent to the temperature. Automatically, this
probability density maximizes a new ``entropy",  and lead us to
pose an equivalent state equation:
\begin{equation}\label{eq:Helmholtz}
\mathcal{ A}= F(\mathbf{q})-S T
\end{equation}
Where $\mathcal{A}$ can be thought as the  corresponding Helmholtz
potential of the equivalent problem. In case of solving the
eigenvalue problem in vibration theory, for example,
$F(\mathbf{q})$ turns to be Rayleigh's quotient $R(\mathbf{q})$
\cite{Newland06,Leissa05}. A briefly mathematical support is given
in appendix \ref{ap:1} which assures  the mathematical possibility
of the probability density proposed.

Summarizing, since our problem of solving the ODE has been
transformed into an equivalent thermal system, all the
mathematical framework of Statistical Mechanics could be applied.
For the sake of brevity, here we only outline how several typical
thermodynamical equilibrium variables such as entropy and specific
heat, could be calculated using previous relations. For example,
for the specific heat $C(T)$ we have, by definition

\begin{equation}\label{eq:calor-especifico}
    C(T)=\frac{d\langle F(\mathbf{q})\rangle }{dT}=\frac{[\langle F(\mathbf{q})\rangle^2-\langle F(\mathbf{q})^2\rangle]}{T^2}
\end{equation}
where $\langle\rangle$ means equilibrium or ensemble average and
could be calculated for each $T$ by Monte-Carlo sampling (see
section \ref{sec:Proposeda}).

Also the entropy could be obtained in the same form applying the
well known relation
\begin{equation}\label{eq:entropia}
   \frac{dS }{dT}=\frac{C(T)}{T}
\end{equation}
which integrated, using equation (\ref{eq:calor-especifico}), give
us
\begin{equation}\label{eq:entropia-intgral}
   S(T)= S(T_1)-\dint_{T_1}^{T} \frac{[\langle F(\mathbf{q})\rangle^2-\langle F(\mathbf{q})^2\rangle]}{T'^3} dT'
\end{equation}
\\

\section{PROPOSED ALGORITHM}\label{sec:Proposeda}
Having established this new equivalent problem, we apply the whole
framework of equilibrium statistical analysis to obtain the
extreme value of $F(\mathbf{q})$. Our main tool for solving
problems through this paper is based on Metropolis Monte-Carlo
algorithm \cite{Metro53}. Briefly, this algorithm ensures a way to
sample canonical ensemble proposing a transition probability
$P_{\mathbf{q',q}}$ which satisfies detailed balance. For our
case, it takes the form
\begin{equation}\label{eq:probtrans}
P_{\mathbf{q',q}}= \frac{\rho_{\mathbf{q'}}}{\rho_{\mathbf{q}}}=
\exp\left(-\frac{\Delta
F\left(\mathbf{q'},\mathbf{q}\right)}{T}\right)
\end{equation}
where $T=1/\beta$, $k_b\equiv 1$.

With this transition probability, we construct an algorithm that
can sample our $n-$\emph{dimensional} configurational phase space
$\mathcal{H}_n$, whose limit as $n\rightarrow \infty$ tends to
Hilbert space $\mathcal{H}_A$. Each equilibrium state, which can
be reached by waiting sufficient time (Monte-Carlo steps), is a
function of parameter $T$ in which all possible configurations
differ from each other in $\pm T$. Obviously, this also means that
for a given $T>0$ there exist many equilibrium configurations and
only for $T=0$ it reduces to one.

The proposed algorithm takes the previous idea and uses it in the
same way as in standard simulated annealing \cite{Kirk83}. We've
just exposed that equilibrium states are governed by parameter
$T$. So, if one diminishes its value in a way so as to go through
successive equilibrium states, then, the most representative
points of the ensemble will be sampled. In the limit as $T
\rightarrow 0$, each different configuration will differ from each
other in an infinitesimal quantity which means, from the notion of
a variation of a functional, that we reach a minimum
$\mathbf{q^{\ast}}=u^{\ast}$ (solution of the ODE). By this way,
the algorithm is supposed to be capable of avoiding relative
minima, providing the absolute minimum of the functional. For the
eigenvalue problem, this allows only to obtain the fundamental
mode and lowest eigenvalue. However, higher modes which correspond
to relative minima, could be obtained in principle if we impose
restrictions to the functional, such as symmetry or number of
nodes; these topics will be explored in future works.

\subsection{Solution generation }
Here, we address the problem of generating candidate solutions
$\mathbf{q'}$ from an existing one $\mathbf{q}$ (solution
generation scheme). To attempt this end, we will first make a
brief survey of the calculus of variations \cite{Hildebrand73}.
The variation of a functional $F(u)=\dint f(u,u^{(1)},...,u^{(s)})
dx$ which depends on the function $u(x)$ and its derivatives of
order $\nu$, $u^{(\nu)}$ is
\begin{eqnarray}\label{eq:var-func}
    \delta F(u,u^{(1)},.,u^{(s)}) &= \epsilon(u) \delta u
\end{eqnarray}

where by definition\\
\begin{displaymath}
\epsilon(u) \delta(u)= \dint \dsum_{\nu=0}^s (-1)^{\nu} \frac{d}{d
x} \left[\frac{\partial f(u,u^{(1)},.,u^{(s)}) }{\partial
u^{(\nu)}}\right]  \delta u dx
\end{displaymath}

 and $\delta u$ is the variation of the function $u(x)$ and is equal to $\epsilon
 \eta(x)$, $\epsilon \rightarrow 0$; also $\eta(x)$ is an arbitrary function
 satisfying sufficiently smooth conditions \cite{Hildebrand73}.

Our solution generation scheme starts generating a Markov chain in
which the next iterate is built from the previous one making the
change $\mathbf{q'}\rightarrow \mathbf{q} + \delta \mathbf{q}$.
The proposal consists in taking $\delta  \mathbf{q}$ in agreement
with the master equation \cite{Chandler87}. We consider $R$ steps
(fixed number of steps) for a given $T$. For the transition
probability to have a value of 1/2 we set  $\Delta
F(\mathbf{q'},\mathbf{q}) \simeq \delta F$; then,
\begin{equation}\label{eq:jumpq}
P_{\mathbf{q',q}}= \frac{1}{2}= \exp\left(-\frac{\Delta
F(\mathbf{q'},\mathbf{q})}{T}\right)\simeq
\exp\left(-\frac{\epsilon(\mathbf{q}) \delta \mathbf{q}
 }{T}\right)
\end{equation}
where equation (\ref{eq:var-func}) has been used. Now, it is
possible to calculate $\delta \mathbf{q}=\{\delta
u_1(x_1),...,\delta u_l(x_l),...\delta u_n(x_n)\}$ considering the
integral in (\ref{eq:var-func}) as a sum over the $n$ sites of the
lattice. According to this, and considering every $\delta
u_l(x_l)$ identical and independent of $x_l$, we can calculate an
estimation of the jump $\delta \overline{u}$ that satisfies
(\ref{eq:jumpq})

\begin{equation}\label{eq:jumpu} \delta u_l(x_l)=\delta
\overline{u}=-\frac{\ln (1/2) T}{\epsilon(u)}
\end{equation}
Then, $\delta \mathbf{q}=\{\delta \overline{u},...,\delta
\overline{u}\}$. This form of calculating $\delta \mathbf{q}$ is
one possible way of setting an estimation to the jump to build the
next iterate. Of course, other possibilities are feasible to be
implemented. However, it has the advantage of efficiently
exploring configurational phase space in order to reach
equilibrium states avoiding local minima (if there exists one).

\subsection{Algorithm implementation}

 It is a relevant problem to determine the initial temperature $T_0$ to start the algorithm.
The one proposed by Kirkpatrick \cite{Kirk83} for initial
temperature guesses in standard simulated annealing technique
(SA), proves to be simple and very effective. Obviously, other
means are possible but we chose it for simplicity (see
\cite{Ceranic01} and references therein). It consists of
conducting a pilot survey of the solution space in which all
increases in the objective function (the functional $F(u)$ ) are
accepted. The initial temperature is then calculated from
(\ref{eq:probtrans}), once the initial transition probability
$P_{\mathbf{q',q_0}}$ has been given ( $0.9\sim0.8$ commonly
selected), according to
\begin{equation}\label{eq:initTemp}
    T_0=-\frac{ \Delta \widetilde{ f}}{\ln(P_{\mathbf{q',q_0}})}
\end{equation}
where $\Delta\widetilde{f}$ is an average increase in the
objective function, $F(u)$. According to this formula it is then
possible to initiate the algorithm. To trigger it for the first
time, a high $T$ is chosen. Then, compute $\Delta \widetilde{f}$
for the first
hundred of attempts and apply (\ref{eq:initTemp}).  \\

In the implementation of the algorithm, we decided to lower the
temperature $T$ half an order of magnitude between different
equilibrium states. Then, we perform $10^5$ runs to compute
different equilibrium averages for a given temperature to
ensure we reach equilibrium.\\

Since it is impossible to reach absolute zero as a limiting value
of the temperature $T$, a stopping criteria must be used as $T
\rightarrow 0$. This criteria could be condensed in the following
way: the process of cooling of the system ends when, making a
choice of a significative digit of the value of the functional
being minimized, this digit averages to zero during the cooling
procedure, while the other, more significative digits don't
change.

\section{THE LANGEVIN APPROACH}\label{sec:Langevina}
 An alternative treatment of the problem is to present it within
 a stochastic framework. To this end, we propose to model the
 system with the Langevin equation.
The use of Langevin equation to stochastically sample an arbitrary
field is not a new idea \cite{Fredrickson02}. However, the authors
ignore that the same approach have been attempted to find
solutions of ODEs.  Briefly,  Langevin equation is a stochastic
 differential equation whose coefficients are random with given
 stochastic properties \cite{VanKampen87}. It defines $u(x,t)$ as a stochastic
 process provided that an initial condition is added, $u(x,0)$.
 The developed method to solve ODEs consists of proposing a
Langevin equation for the scalar field $u(x)$ in the simulation
time $t$, this is
\begin{equation}\label{eq:Langevin}
    \frac{\partial u(x,t) }{\partial t}=-1\frac{1}{\xi(T)} \frac{\delta F[u]}{ \delta
    u}+ \eta(x,t)
\end{equation}
where $\eta(x,t)$ is a Gaussian thermal noise with zero mean and
that satisfies
\begin{equation}\label{eq:noise}
<\eta(x,t) \eta(x',t')>=2D(T)\delta(x-x') \delta(t-t')
\end{equation}
There is a certain arbitrariness on the selection of $D(T)$ and
$\xi(T)$ but the product $\xi(T) D(T)$ must satisfy Einstein
relation, i.e. $\xi(T) D(T)=k_b T$ to give the correct
thermodynamical averages. It can be proved that the probability
distribution $P(u,t)$ approaches the equilibrium distribution
$P_{eq}\propto e^{-F(u)/k_bT}$ as $t\rightarrow \infty$
\cite{Batrouni87}. To obtain the minimum of $F(u)$ and
consequently the solution of the ODE, we must solve
(\ref{eq:Langevin}) starting at a high temperature $T_0$ and then
slowing it down gradually, which means passing through successive
equilibria, until we reach a temperature near zero. Following the
same reasoning as before we will obtain, by this way, the absolute
minimum of the functional. Of course, the implementation of this
cooling schedule is analogous to the previously developed
procedure, so there is no need to go into detail again.

In the end, to make possible a numerical solution of the field
equation (\ref{eq:Langevin}), we work with the same discretization
scheme as above. This time,  Langevin equation for the scalar
lattice field $u_l(x_l)$ looks like
\begin{equation}\label{eq:Langevindiscrete}
    \frac{\partial u_l(x_l,t) }{\partial t}=-1\frac{1}{\xi(T)} \frac{\delta F[u_l]}{ \delta
    u_l(x_l)}+ \eta(x_l,t) \hspace{0.2in} l=1..n
\end{equation}

\section{NUMERICAL SIMULATIONS}\label{sec:Numericals}
In this section we present several ODEs which were solved applying
the previously developed methods. The cases under study are three

\begin{itemize}
    \item Case 1: Helmholtz equation.
    \item Case 2: Fourth order beam equation.
    \item Case 3: Fourth order beam  deflection under static load.
\end{itemize}

The proposed methods provide, for the eigenvalue problem (cases 1
and 2), the absolute minimum of the functional (lowest eigenvalue
) and the corresponding eigenfunction (fundamental mode). For case
3, this situation no longer exists since there is only one minimum
of the functional and this represents the unique solution of the
ODE.

\subsection{Case 1}
For the case of Helmholtz equation, $u''(x)+ \lambda u(x)=0$, the
corresponding functional $F(u)$ reduces to

\begin{equation}\label{eq:functHelmholtz}
    F(u)=\frac{\dint_0^L (u'(x))^2 dx}{\dint_0^L u(x)^2 dx}
\end{equation}
 This result is obtained from equation (\ref{eq:eigen-funct}) after an
easy manipulation. Fixed boundary conditions will be considered in
this case, this means $u(0)=u(L)=0$. The functional is named
Rayleigh quotient $R(\mathbf{q})$, as said before. Physically, it
can be derived from the principle of conservation of energy since,
if the total energy of the system remains constant, one can assure
that $V_{max}=T_{max}$.

For a vibrating system, let $T_{max}=\omega^2T^*_{max}$; that is,
$T^*_{max}$ is the maximum kinetic energy of the system during a
cycle of motion, with the square of the natural frequency,
$\omega^2$, factored out. Thus, from the total constant energy
requirement, one can write

\begin{equation}\label{eq:eq:Rayleigh}
   \omega^2= \frac{V_{max}}{T^*_{max}}
\end{equation}
which is exactly (\ref{eq:functHelmholtz}) with
$\omega^2=\lambda$.

 For the algorithm implementation, we start with a random function $\mathbf{q_0}$
and then applies the solution generation routine and the
transition probability $P_{\mathbf{q',q}}$ of (\ref{eq:probtrans})
together with the cooling schedule till the final temperature
criterion is reached. For this case we have discretized the 1
dimensional lattice in 16 sites, i. e.  $n=16$. However, to
compute the functional values $F(u)$, we have employed a cubic
spline interpolation scheme using $Matlab^{R}$ routines
between $u_l$'s values which smooths the representation of $u(x)$ to calculate $F(u)$.\\
For the algorithm to perform the iterations over different
configurations, new configurations are settled from old ones with
a jump estimation $\delta \mathbf{q}=\{\delta
\overline{u},...,\delta \overline{u}\}$ that is calculated from
equation (\ref{eq:jumpu})

\begin{equation}\label{eq:deltauHelm}
 \delta \overline{u}=-\frac{ \ln(1/2) T ||u||^2}{2 G(u)}
\end{equation}
where $G(u)=\int (u''+\lambda u) dx$ represents Galerkin's error
function and $||u||^2=\int u^2 dx$. Here $\epsilon(u)=2
G(u)/||u||^2$ as can be seen by comparing equations
(\ref{eq:deltauHelm}) and (\ref{eq:jumpu}). Obviously, $u''(x)+
\lambda u(x)=0$ is only satisfied by the solution of the ODE
$u^{\ast}$.

\begin{figure}[h]
\centerline{\includegraphics[scale=0.6]{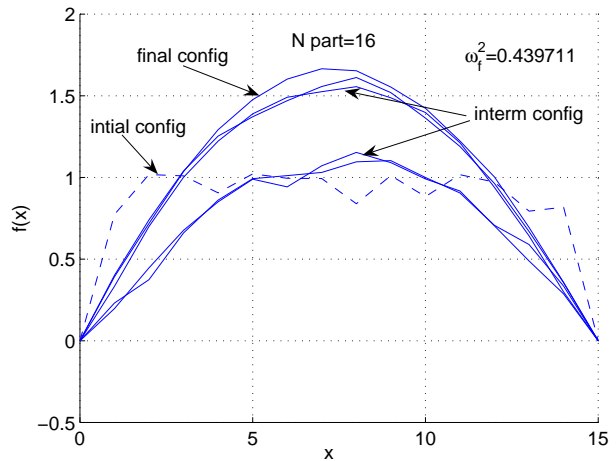}} \caption{Initial,
intermediate and final configurations for the Helmholtz equation
with fixed ends, showing the convergence of the algorithm to the
minimum of the functional, $N part=16$.}\label{fig:2}
\end{figure}

The initial random function $\mathbf{q_0}$ can be observed in
figure (\ref{fig:2}) as \emph{initial config}. In the same figure,
intermediate candidate solutions which correspond to different
\emph{equilibrium} situations for different temperatures are
shown. The final configuration was chosen following the final
temperature stopping criterium. Numerical values of the considered
configurations are presented in table (\ref{tab:fixedHelm}) where
the number of algorithm iterations is also shown for time's
estimation calculations.

\begin{table}[htb]
\centering
\begin{tabular}{ c |c c |c }
  \emph{config }& \emph{R(u)}& $T$ &\emph{N iter}  \\
\hline
       initial & $0.396959$ & $1\cdot 10^{-4}$  &   0        \\
\hline
      interm 1 & $0.083343$ & $1\cdot 10^{-4}$  &  $1\cdot 10^{3}$   \\

      interm 2 & $0.048746$ & $1\cdot 10^{-4}$  &   $2\cdot 10^{3}$    \\

      interm 3 & $0.045338$ & $1\cdot 10^{-4} $ &  $3\cdot 10^{3}$    \\
\hline
      final 16 & $0.043871$ & $1\cdot 10^{-6}$  &  $ 1\cdot 10^{4}$    \\

\end{tabular}
\caption{Rayleigh's quotient for the configurations of figure
\ref{fig:2} with $N part=16$. Temperature  and number of algorithm
iterations are also shown.} \label{tab:fixedHelm}
\end{table}

\begin{figure}[h]
\centerline{\includegraphics[scale=0.6]{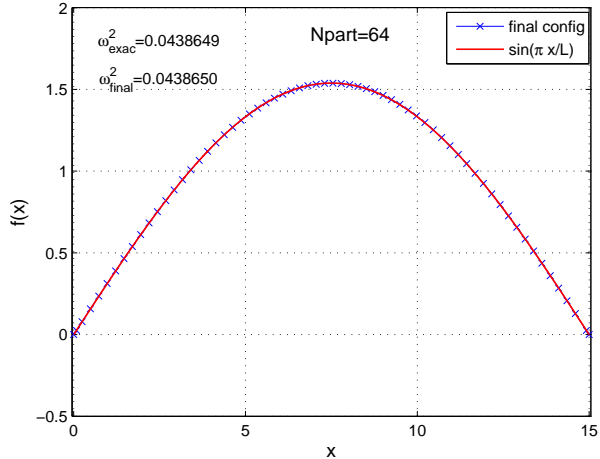}} \caption{Final
configuration $\times$ (cross) and exact solution $-$ (full line)
for the Helmholtz equation with  $N part=64$.}\label{fig:3}
\end{figure}

The selection of $n=16$ or hereafter $N part=16$ was not
arbitrary. A few number of particles at the beginning of the run
makes the algorithm to run faster at large values of the
prescribed $F(u)$ (we can call it the ``energy" of the system). As
this ``energy" goes down, the algorithm starts to slow down its
convergence rate due to the finite number of particles. At this
stage, we increment the number of particles as many times as it is
necessary to get further significant reductions. This process has
been named ``stable resizing" due to the ``resize" of the number
of considered particles. Figure (\ref{fig:3}) shows the results
for $N part=64$. The exact solution (fundamental mode) appear in
full line and the function that results of the minimization
process is shown in cross line. The numerical values of Rayleigh
quotient are presented in table (\ref{tab:fixedHelmfinal}). It can
be noted that the difference between the exact and obtained $R(u)$
is below 0.01\% which demonstrates the convergence of the method
to the exact solution.

\begin{table}[htb]
\centering
\begin{tabular}{ c |c c |c}
  \emph{config} & \emph{R(u)} & $T$  & \emph{N iter}  \\
\hline
       final 64 & $0.0438650$ & $1\cdot 10^{-9}$  &  $ 5\cdot 10^{4}$   \\
       exact    & $0.0438649$ & -- &   --          \\
\end{tabular}
\caption{ Rayleigh's quotient for the final configuration of
figure \ref{fig:3} with $N part=64$. Here, the difference between
the exact solution and the solution provided by the algorithm
(final configuration) differs only in $0.01 \%$.
}\label{tab:fixedHelmfinal}
\end{table}

\subsection{Langevin approach to the Helmholtz equation}
For the case of the Helmholtz equation $u''(x)+ \lambda u(x)=0$,
we have already shown that the corresponding functional is
$F(u)=\int_0^L (u'(x))^2 dx/\int_0^L u(x)^2 dx$ which, in its
discretized version is
\begin{equation}\label{langevin-helmholtz}
    F(u_l)=\dsum_{l=1}^{n} (\nabla u_l)^2 \delta x_l/\dsum_{l=1}^{n} u_l^2 \delta x_l
\end{equation}
where $\nabla u_l=(u_{l+1}-u_l)/\delta x_l$. With this form of
$F(u_l)$ and applying an Euler updating scheme \cite{Batrouni87},
it can be substituted in (\ref{eq:Langevindiscrete}) to give

\begin{eqnarray*}\label{langevin-helmholtz2}
  u_l(t+\delta t)=u_l(t)- \delta t \left(-2\frac{(u_{l+1}-2u_{l}+ u_{l-1})/\delta x_l + \lambda u_l \delta x_l}{\|u_l\|^2}\right)\\
  + \sqrt{\delta t} \sqrt{ 2 D(T)}\eta_l(t)
\end{eqnarray*}
\\where
\begin{displaymath}
\|u_l\|^2=\dsum_{l=1}^{n} u_l^2 \delta x_l
\end{displaymath}

\begin{figure}[htb]
\centerline{\includegraphics[scale=0.4]{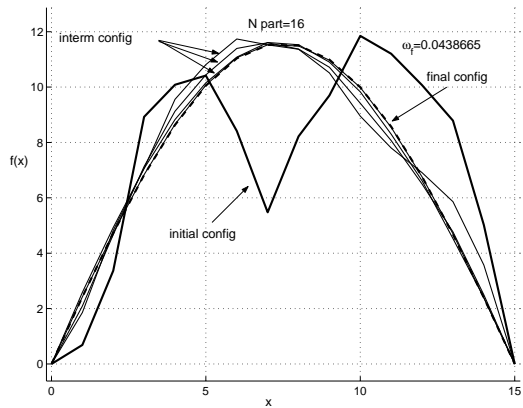}}
\caption{Initial, intermediate and final configurations for the
Helmholtz equation with fixed ends for the simulations with the
Langevin approach, showing the convergence of the algorithm to the
minimum of the functional, $N part=16$.}\label{fig:Langevin16}
\end{figure}

Fixed boundary conditions are also considered in this case.
Numerical values of the obtained configurations are presented in
table (\ref{tab:fixedHelmLang}) where the number of algorithm
iterations is also shown for time's estimation calculations.

\begin{table}[htb]
\centering
\begin{tabular}{ c |c c |c }
  \emph{config }& $R(u)$& $T$        &\emph{N iter}  \\
\hline
       initial & 0.273213   & --  &   0        \\
\hline
      interm 1 & 0.115614 & $1\cdot 10^{-2}$  &  $1\cdot 10^{4}$   \\

      interm 2 & 0.053995 & $1\cdot 10^{-3}$  &  $1\cdot 10^{4}$    \\

      interm 3 & 0.044929 & $1\cdot 10^{-4}$  &  $1\cdot 10^{4}$    \\

      interm 4 & 0.044005 & $1\cdot 10^{-5}$  &  $1\cdot 10^{4}$   \\

      interm 5 & 0.043875 & $1\cdot 10^{-6}$  &  $1\cdot 10^{4}$    \\

\hline
      final 16 & 0.043866 & $1\cdot 10^{-7}$  &  $ 1\cdot 10^{4} $   \\

\end{tabular}
\caption{Rayleigh's quotient for the configurations of figure
\ref{fig:Langevin16} with $N part=16$ for simulations with
Langevin approach. Temperature $T$ and number of algorithm
iterations are also shown.} \label{tab:fixedHelmLang}
\end{table}

A stable resizing routine could also be applied in this case. The
results for $N part=64$ are shown in table
(\ref{tab:fixedHelmLangfinal}).

\begin{table}[htb]
\centering
\begin{tabular}{ c |c c |c}
  \emph{config} & $R(u)$& $T$  & \emph{N iter}  \\
\hline \hline
       final 64 &0.0438661 & $1\cdot 10^{-8}$  &   $1\cdot 10^{4}$   \\
       exact    &0.0438649 & -- &   --          \\
\end{tabular}
\caption{ Rayleigh's quotient for the final configuration of the
simulations with Langevin approach with $N part=64$; the exact
value of Rayleigh's quotient is shown for comparison.
}\label{tab:fixedHelmLangfinal}
\end{table}

\subsection{Case 2}
For the fourth order beam equation, $u^{(IV)}(x)- \lambda u(x)=0$,
the corresponding functional is
\begin{equation}\label{eq:functbeam}
    F(u)=\frac{\dint_0^L (u''(x))^2 dx}{\dint_0^L u(x)^2 dx}
\end{equation}

Clamped boundary conditions are considered for this case, i. e.
$u(0)=u(L)=\partial u(x)/\partial x|_{x=0}=\partial u(x)/\partial
x|_{x=L}=0$. As it was done before, the jump estimation can be
calculated from equation (\ref{eq:jumpu}) resulting in the same
expression as (\ref{eq:deltauHelm}). The only change is in
$\epsilon(u)=2 G(u)/||u||^2$. This time, $G(u)$ is
\begin{equation}\label{eq:epsilonbeam}
 G(u)=\int (u^{(IV)}-\lambda u) dx
\end{equation}

\begin{figure}[htb]
\centerline{\includegraphics[scale=0.5]{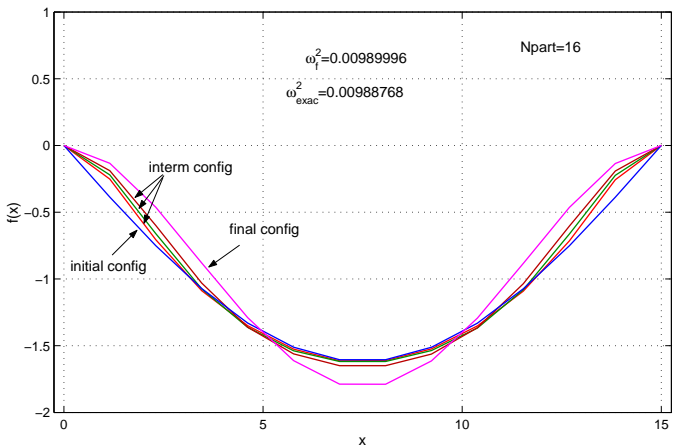}} \caption{Initial,
final and intermediate configurations as the algorithm runs
towards the solution of the beam equation with clamped ends. $N
part=16$.}\label{fig:6}
\end{figure}

We start the algorithm with a sinusoidal function, $\mathbf{q_0}$,
which is not the solution of the differential equation in this
case (see Figure \ref{fig:6}). This was made to illustrate that
the algorithm converges to the solution of the ODE independently
of the initial condition that has been employed. In figure
(\ref{fig:6}) we also show the convergence of the algorithm to the
solution of the beam equation as temperature $T$ goes down. The
number of particles is $N part=16$ for the first step of the
process as was explained before. Numerical values are shown in
table (\ref{tab:sinbeam}).

\begin{table}[htb]
\centering
\begin{tabular}{ c|c c|c }
 \emph{ config} & \emph{R(u)}& $T$ & \emph{N iter}  \\
\hline
       initial & 0.0372191 & $5\cdot 10^{-6}$   &   0             \\
      interm 1 & 0.0168120 & $1\cdot 10^{-6}$   &  $2\cdot 10^{4}$   \\
      interm 2 & 0.0144376 & $7.5\cdot 10^{-7}$ &  $4\cdot 10^{4} $  \\
      interm 3 & 0.0123598 & $2.5\cdot 10^{-4}$ &  $6\cdot 10^{4}$   \\
\hline
      final 16 & 0.0098999 & $1\cdot 10^{-9}$   &  $ 2\cdot 10^{5}$  \\
\end{tabular}
\caption{Rayleigh quotient for the configurations shown in figure
\ref{fig:6} (fourth order beam equation);  $N part=16$.
Temperature $T$ and the number of iterations are also shown.}
\label{tab:sinbeam}
\end{table}

\begin{figure}[htb]
\centerline{\includegraphics[scale=0.6]{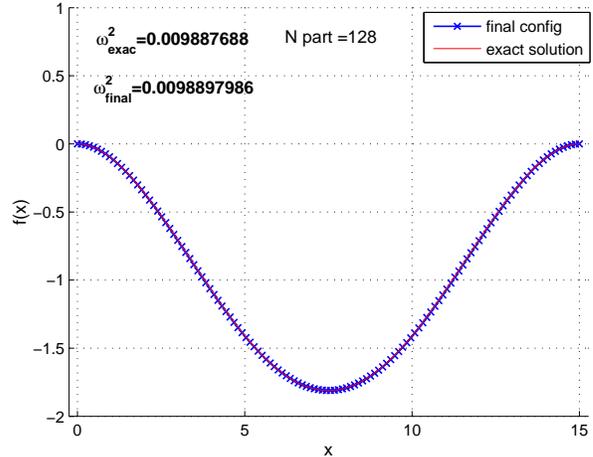}} \caption{Final
configuration $\times$ (cross) and exact solution $-$ (full line)
of a beam with clamped ends; $N part=128$.}\label{fig:7}
\end{figure}

To improve the precision of the algorithm and accelerate its
convergence, the number of particles were increased five times ($N
part=128$) with respect to its initial value. The final results
can be observed in figure (\ref{fig:7}) where the Rayleigh
quotient $R$ is also shown. It is noticeable the better accuracy
of the attained solution due to the ``stable resizing".

\subsection{Case 3}
For the fourth order beam deflection under a static load the
equation is $u^{(IV)}(x)=g(x)$. We have selected a  concentrated
load of the form $g(x)=\delta(x-x_a)$ being $\delta(x)$, the Dirac
delta function. This time, the functional is
\begin{equation}\label{eq;functbeamdeflect}
F(u)=\int_0^L (u''(x))^2 dx - 2\int_0^L u(x) g(x) dx
\end{equation}

Simply supported boundary conditions are selected this time,
$u(0)=u(L)=\partial^2 u(x)/\partial x^2|_{x=0}=\partial^2
u(x)/\partial x^2|_{x=L}=0$. In this case, the functional simply
represents twice the potential energy in a beam subjected to the
transverse force $g(x)$. Applying equation (\ref{eq:jumpq}) one is
able to obtain the estimated jump which results
\begin{equation}\label{eq:jump-staticbeam}
\delta \overline{u}=\frac{- \ln(1/2) T + 2 \delta u(x_a)}{2 G(u)}
\end{equation}
where $G (u)= \int u^{(IV)} dx$. This equation states that the
estimated jump could be obtained once  $\delta u(x_a)$ is
provided. Since we have no knowledge of this quantity, it is
reasonable to propose to be of order of $\delta \overline{u}$.
With this assumption, equation (\ref{eq:jump-staticbeam}) results
\begin{equation}\label{eq:jump-staticbeam2}
\delta \overline{u}=-\frac{\ln(1/2) T}{2 (G(u)-1)}
\end{equation}

\begin{figure}[htb]
\centerline{\includegraphics[scale=0.6]{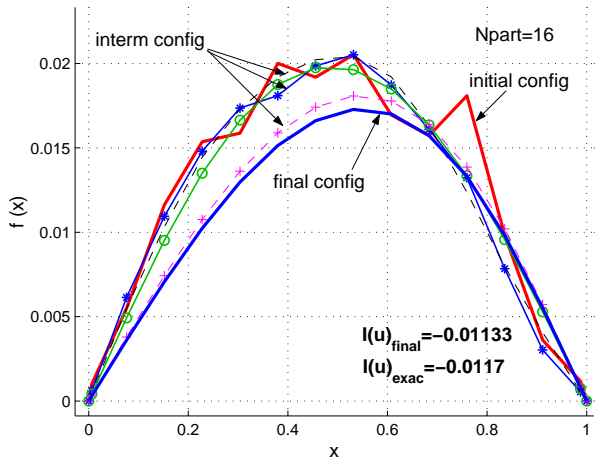}} \caption{Initial,
final and intermediate configurations in the convergence to the
solution of a beam under static load; $N part=16$.}
 \label{fig:9}
\end{figure}

The algorithm was initialized with an initial configuration of the
form of a sine function with random noise. This configuration,
together with intermediates and final configurations are shown in
figure (\ref{fig:9}). $N part=16$ was chosen again since it
provides good initial results. Numerical values of the different
configurations are shown by table (\ref{tab:beamest}) for
$x_a=0.5L$, where we rename $I(u)\equiv F(u)$ .

\begin{table}[htb]
\centering
\begin{tabular}{ c| c c |c }
   \emph{config} & $I(u)$ & $T$   & $N iter$  \\
  \hline
   initial & 1.7231 & $1\cdot 10^{-4}$    & 0  \\

   interm 1 &0.1172 & $1\cdot 10^{-4}$    & $1\cdot10^{3}$ \\

   interm 2 &0.001995 & $1\cdot 10^{-4}$  & $5\cdot 10^{3}$  \\

    interm 3&-0.0038  & $1\cdot 10^{-4}$   & $1\cdot 10^{4}$  \\

    interm 4&-0.00808 & $1\cdot 10^{-6}$ & $ 2\cdot 10^{4}$  \\

    interm 5&-0.01103 & $1\cdot 10^{-7}$   & $1\cdot 10^{5}$ \\
\hline
    final   & -0.011334 & $1\cdot 10^{-8}$  & $2\cdot 10^{5}$\\

\end{tabular}
\caption{Functional value, $I(u)$, for the case of the elastic
beam equation under a static force of delta type with simply
supported boundary conditions.} \label{tab:beamest}
\end{table}

If a better precision wants to be reached, more particles must be
included. Nevertheless with $N part=16$ the difference between the
exact and final solution is below 3\%.

\section{CONCLUSIONS}
A statistical approximation to the solution of ODEs were
presented, in particular we solved some problems related with
vibration theory. The developed method establishes a way to attain
the solution of an ODE proposed to solve physical problems, from
statistical mechanics. The developed idea consists of thinking the
domain of any functional (representing the ODE) as microstates of
a thermal system (particle system) in contact with a heat bath at
temperature $T$, that interacts trough a thermodynamic potential
represented by the functional. Then, obtaining the minimum of the
thermodynamic potential of this equivalent problem when
$T\rightarrow 0$ is the same as to get the solution of the ODE. If
only one solution exists, our method provides the unique solution
of the ODE. In case we treat an eigenvalue equation, where
infinite solutions exist, we obtain the absolute minimum of the
corresponding functional or fundamental mode, as well as the
fundamental eigenvalue.

Three examples have been provided to show the convergence of
method to the solution of the ODEs; the first two of them,
represents a typical eigenvalue problem, as it is the case for the
Helmholtz equation and the fourth order beam equation, with
infinite possible solutions. In these cases the method have shown
to converge to the fundamental mode. For the third case, the beam
equation under a static load, the method converge to its unique
solution. Additionally, a resizing of the corresponding one
dimensional lattice called ``stable rezising", which increases the
discretization, was used to provide further reductions of the
functional accelerating its convergence and improving its
precision.

In the same spirit, a general approach to the problem within an
stochastic framework was presented with similar success. We
proposed an equivalent Langevin equation for the scalar field $u$
 in its way to find the minimum of the functional, solution of the
ODE. Numerical experiments have been run that showed that the
method also converges to the exact solution for the Helmholtz
equation. Of course, the other two cases, analyzed with the first
developed method, can be treated within the same scheme but, for
the sake of brevity, we postpone them for future works.

Summarizing, the present article does not pretend to claim a place
as an alternative method to other numerical, probably faster and
more precise, methods. Its virtue consists in that it is possible
to establish a link between the problem of finding the solution of
ODEs and an interacting particle system which can be treated by
statistical mechanics; in this sense, we showed that some typical
thermodynamical equilibrium variables, such as entropy and
specific heat, could be in principle calculated. This provides a
different perspective and new insights in solving ODEs.

\section{Acknowledgments} M. F. and S. A. V.  were supported by
CONICET (Argentina) and by Secretar\'ia General de Ciencia y
Tecnolog\'ia of Universidad Nacional del Sur at the Department of
Physics.

\appendix
\section{}\label{ap:1}

When modelling a thermal system in equilibrium, the  supportive
mathematical  structure is simple and not exclusive of physical
systems. In this mathematical structure, the entropy $S$ is
defined over a convex subset $\sum$ of $\mathbf{R}^{m+1}$, where a
point of $\mathbf{R}^{m+1}$ is denoted by
$(X_{0},X_{1}....X_{m})$ so it can be defined a $C^{1}$ function\\
 $ S : \Sigma \rightarrow \mathbf{R}$\\
such that \cite{Evans02}:
\begin{itemize}
    \item $ S$ is concave
    \item $\frac{\partial S}{\partial E}>0$, being $E=X_{0}$
    \item $S$ is positively homogeneous of degree 1
\end{itemize}
For a probabilistic model there is also a measure needed, then,
the requirement is fulfilled if:
\begin{itemize}
    \item there is a class $\mathcal{A}$ defined by $\mathcal{A}=\{\rho: \; \Omega
    \rightarrow (0,\infty)\}$ ( in physical systems, a point of $ \Omega$ is called a
    microstate)
    \item  $\rho$ is $\pi-measurable$, ($\pi$ is the reference measure)
    \item $\int_{\Omega}\rho \;d\pi\,
=1$, ( $\rho$ is the density of the microstate measure $\rho
d\pi$)
    \item $E(X,\rho)=<\mathbf{X}>=\int_{\Omega} \mathbf{X}\rho
\;d\pi$, ($X$ is called an observable).
\end{itemize}

There is a well known theorem \cite{Evans02} which ensures that
 \begin{equation}\label{sigma}
\sigma = \frac{\exp{(-\beta \cdot X)}}{Z}
\end{equation}
 maximizes $S(\rho)$ for all possible $\rho  \in \mathcal{A}$. Here, $\beta \in \mathbf{R}^{m+1}$
and
\begin{equation}\label{eq:Z}
 Z = \int_{\Omega}e^{-\beta \cdot X} \, d\pi
\end{equation}
\appendix


\begin{thebibliography}{1}

\bibitem{Hildebrand73} F. Hildebrand, 1973. Methods of applied mathematics.
Eudeba University Press.

\bibitem{Kaplan85} W. Kaplan, 1985. Advanced Calculus. CECSA Editors.

\bibitem{Rektorys75} K. Rektorys, 1975. Variational methods in
mathematics, physics and engineering. D. Reidel Publishing Co.

\bibitem{Chandler87} D. Chandler, 1987. Introduction to modern
statistical mechanics. Oxford University Press.

\bibitem{Sommerfeld56} A. Sommerfeld, 1956. Thermodynamics and statistical mechanics. Academic Press.

\bibitem{Evans02} L. C. Evans, 2002. Entropy and partial
differential equations. Dept. Math. Berkeley.

\bibitem{Reichl98}L. E. Reichl, 1998. A Modern Course in Statistical Physics, 2nd
Edition, Wiley.

\bibitem{Kirk83} S. Kirkpatrick, C. D. Gelatt, M. P. Vecchi, 1983.
Optimization by simulated annealing. Science 220 (4598), 671-680.

\bibitem{Batrouni87} G. G. Batrouni, B. Svetitsky, 1987. Accelerated
dynamics in simulations of first-order phase transitions. Phys.
Rev. B. 36,10 pp5647-5650.

\bibitem{Newland06} D. E. Newland, 2006.
\emph{Mechanical Vibration Analysis and Computation}, Dover,
Mineola, NY, USA.

\bibitem{Leissa05} A. W. Leissa, 2005. The historical basis of the
Rayleigh and Ritz methods. Journal of Sound and Vibration 287 pp
961-978.

\bibitem{Metro53} A. Metropolis, W. Rosenbluth, M. N. Rosenbluth,
H. Teller, E. Teller 1953. Equation of static calculations by fast
computing machines. Journal of Chemical Physics 21 (6), 1087-1092.


\bibitem{Ceranic01} B. Ceranic, C. Fryer, R. W. Baines, 2001.
 An application of simulated annealing to the optimum design of
reinforced retaining structures. Comp. and Struc. 79, pp1569-1581.


\bibitem{Fredrickson02}G. H. Fredrickson, V. Ganesan, F. Drolet, 2002.
Field-Theoretic Computer Simulation Methods for Polymers and
Complex Fluids. Macromolecules 35, pp 16-39.



\bibitem{VanKampen87} N. G. Van Kampen. 1987,
\emph{Stochastic Processes in Physics and Chemistry},
North-Holland, Amsterdam, The Netherlands.


\bibitem{NumRec88} Numerical Recipes in C: The art of scientific
computing, 1988. Cambridge University Press.

\end{thebibliography}
\end{document}